\documentclass[11pt,a4paper]{article}
\usepackage[T1]{fontenc}
\usepackage{lmodern,microtype,amsmath,amssymb,booktabs}
\usepackage[a4paper,margin=28mm]{geometry}
\usepackage[hidelinks]{hyperref}
\newcommand{\lead}[1]{\textbf{#1.}}
\newcommand{\caus}{\rho_{\mathrm{late}}}
\newcommand{\know}{\rho_{\mathrm{early}}}
\newcommand{\EE}{\mathcal E}
\newcommand{\supports}{\mathcal A_k}
\newcommand{\arx}[1]{\href{https://arxiv.org/abs/#1}{arXiv:#1}}
\newcommand{\bref}[3]{\bibitem{ref#1} #2. #3.}
\hypersetup{pdftitle={The Privacy Cost of Writing an Audit Receipt Before Its Context Is Known}}
\title{The Privacy Cost of Writing an Audit Receipt\\Before Its Context Is Known}
\author{Faruk Alpay\thanks{Correspondence: \href{mailto:alpay@lightcap.ai}{\texttt{alpay@lightcap.ai}}.}\\
\small Department of Computer Engineering\\
\small Bah\c{c}e\c{s}ehir University\\
\small Istanbul, T\"urkiye\\
\small \href{mailto:faruk.alpay@bahcesehir.edu.tr}{\texttt{faruk.alpay@bahcesehir.edu.tr}}
\and
Taylan Alpay\\
\small Department of Aerospace Engineering\\
\small University of Turkish Aeronautical Association\\
\small Ankara, T\"urkiye\\
\small \href{mailto:s220112602@stu.thk.edu.tr}{\texttt{s220112602@stu.thk.edu.tr}}}
\date{}
\hypersetup{pdfauthor={Faruk Alpay and Taylan Alpay}}
\begin{document}
\maketitle
\begin{abstract}
An audit recorder must permit exact recovery with a small private index before it knows which questions a future reader will ask. A recursion on question partitions generates the decision trees, their eventwise costs and the price equations. A common-optimal-tree criterion explains when reduction to a fractional cover preserves the result. On uniform binary linear event spaces with four private roles, if every quartet is split evenly by some bit, optimal identification probability at expected follow-up budget $b$ is $(1+b)/2$. Otherwise it is $\min(1/2+b/3,3/4)$. We characterize all priors attaining the latter curve in its rising regime and derive a closed-form compiler for three binary fields. For a population tilted by one pairwise parity we obtain the full budget curve and a single recorder attaining it. On eight negotiated TLS suites the aligned populations allow a 60 percent follow-up rate at a 70 percent identification ceiling. A pairwise correlation with unchanged marginal field frequencies reduces that maximum to 40 percent. Across 960 local TLS sessions, the recorder recovered every negotiated suite using two private bits per session.
\end{abstract}
\section{Introduction}
A reviewer must be able to recover an event's exact identity. The recorder can retain a small confidential index and write a public receipt. Later a review discloses a class containing the event. For instance, an archive could disclose which policy covered a cryptographic operation. Several policies may cover the same operation. The question is how much extra guessing power the permanent receipt gives a reader who learns the policy.
\par Consider three facts $a,b,c$. A future review will identify one of the three pairs that contains the true fact. One private bit suffices to index either member of a known pair. When the pair is learned later, perfect privacy would require the receipt laws of $a$ and $b$ to coincide, and likewise for the other pairs. All three laws must then coincide. Any receipt that can occur must allow the reviewer to recover three possible facts from one bit. This is impossible. The optimum guessing factor is exactly $3/2$ (Theorem~2).
\par Two different timings arise in practice. Sections~2--10 study a policy determined independently of the receipt conditional on the event. Section~\ref{sec:feedback} gives the reader the right to choose its next question after seeing the answer history. Section~\ref{sec:codes} solves the entire budget curve on uniform binary linear event spaces with three or four private roles. Its TLS experiment shows when traffic correlations force the reviewer to reduce its quota.
\subsection{Security experiment}
Let $V$ be a finite set of possible facts and $\EE$ a finite family of nonempty subsets covering $V$. A valid pair $(x,e)$ satisfies $x\in e$. At a single checkpoint the trusted recorder emits a finite public receipt $R$ and keeps $S\in[k]$, where $1\le k\le |V|$. A trusted reviewer must recover $x$ from $(R,S)$ with zero error on every legal pair. In the late experiment the first encoder sees only $x$. In the early experiment it sees $(x,e)$. In both games the archive observer sees $(R,e)$ and can retain it indefinitely. The decoder can be fixed independently of $e$ in both constructions. A private state of cardinality $k$ represents $k$ distinguishable values, which requires $\lceil\log_2 k\rceil$ bits in a fixed-width implementation.
\section{The exact receipt optimization}
Contexts are exogenous conditional on the fact. In the late game their selection is independent of the recorder's private coins given $x$. No auxiliary secret key, side archive, uncharged decoder memory, or private transcript is available to the reviewer. An observer who already knows $e$ scores the additional information in $R$. For a late channel $W(r\mid x)$ its worst-context guessing factor is
\begin{equation}
  \rho(W)=\max_{e\in\EE}\sum_r\max_{x\in e} W(r\mid x). \label{eq:metric}
\end{equation}
Its logarithm in base two is conditional maximal leakage [1,2] for any input law with full support on the incidence relation. For a uniformly drawn fact inside a fixed context, its expression before the outer maximum is exactly the ratio of optimal guessing probabilities with and without the receipt. The score is evaluated after disclosure. It allows an early encoder to reveal the label itself in advance. A requirement to hide the label until announcement would change the comparison. The early game uses a separate channel $W_e$ on each fiber. Let $\caus$ and $\know$ be the corresponding minimum factors.
\par\lead{Theorem 1} With $\supports=\{A\subseteq V:1\le |A|\le k\}$, the two optima are
\begin{align}
\caus&=\min_{w_A\ge0}\ \max_{e\in\EE}\ \sum_{A\cap e\ne\varnothing} w_A,
&\quad \sum_{A\ni x}w_A=1\quad (x\in V),\label{eq:lp}\\
\know&=\max_{e\in\EE}\max\{1,|e|/k\}. \label{eq:early}
\end{align}
All sums over $A$ in (\ref{eq:lp}) use $\supports$. An epigraph variable for the maximum gives a linear program with rational data. The optimum is attained.
\par\lead{Proof} Fix a late channel. For any $r$, at most $k$ facts can produce it, since the decoder has only $k$ possible outputs at that $r$. Conversely, any channel with these column supports permits exact recovery by privately retaining the index within the support. Write a column as nested indicators at its distinct positive levels. More precisely, set $0=u_0<u_1<\cdots<u_m$ to its levels and $A_j=\{x:W(r\mid x)\ge u_j\}$. Add $u_j-u_{j-1}$ to $w_{A_j}$. This preserves the row sum at every $x$ and, simultaneously for every $e$, the column maximum on $e$. Summing over receipts gives a feasible cover of equal cost. Conversely, given $w$, emit $A$ with probability $w_A$ whenever $x\in A$ and store the rank of $x$ in a fixed ordering of $A$. Its score is exactly (\ref{eq:lp}).
\par For one known context of size $m$, row normalization and the support bound give $m\le k\sum_r\max_x W_e(r\mid x)$, and the factor is always at least one. If $m\le k$, emit the context itself and retain the rank. Otherwise choose a uniform $k$-subset of the context containing $x$ and retain its rank. The factor is $\binom{m}{k}/\binom{m-1}{k-1}=m/k$. Conditioning the construction on each $e$ proves (\ref{eq:early}).\hfill$\square$
\par\lead{Corollary 1} Join two facts if they share a context. Then $\caus=1$ exactly when each connected component has size at most $k$. Indeed, a factor of one on a context forces all its channel rows to be identical by termwise equality with their coordinatewise maximum. Equality propagates along paths. Each common nonempty column therefore supports its whole component. Conversely, a receipt may name the component and a private index may name the fact within it. The context already identifies the component.
\section{When delay has an unbounded cost}
Write $\Gamma(A)=\{e\in\EE:A\cap e\ne\varnothing\}$. Suppose a group of permutations of $V$ preserves the incidence relation and is transitive separately on vertices and on contexts. Let $n=|V|$, $m=|\EE|$. Then
\begin{equation}
 \caus=\frac{n}{m}\min_{A\in\supports}\frac{|\Gamma(A)|}{|A|}.\label{eq:exp}
\end{equation}
To see the lower bound, average the objective of (\ref{eq:lp}) over contexts and use $\sum_A |A|w_A=n$. For the reverse bound, choose a minimizing $A$, give all sets in its orbit equal weight, and normalize at one vertex. Vertex transitivity gives the same row sum at every vertex. Context transitivity gives the same cost at every context, so the average lower bound is attained. The proof also applies to a permutation-invariant restriction on allowed supports that is closed under taking nonempty subsets.
\par\lead{Theorem 2} If $\EE$ consists of all $d$-subsets of $[n]$, then
\begin{equation}
 \caus=\frac{\binom nk-\binom{n-d}k}{\binom{n-1}{k-1}},\qquad
 \know=\max\{1,d/k\}. \label{eq:complete}
\end{equation}
Here $\binom ab=0$ for $b>a$. To verify optimality at size $k$, the fraction of $d$-contexts hit by a fixed $s$-set is
$f(s)=1-\binom{n-s}{d}/\binom nd$. Its successive increments are $\binom{n-s-1}{d-1}/\binom nd$ for $s=0,\ldots,n-1$, with the zero convention. Hence $f(s)/s$ is nonincreasing. Substitution at $k$ in (\ref{eq:exp}) gives (\ref{eq:complete}). Uniformly drawing a $k$-subset containing $x$ is an optimal causal encoder.
\par\lead{A growing separation} Set $n=q^2$, $k=d=q$. With early context the leakage is zero. With late context,
\[
 \caus=q\left(1-\prod_{j=0}^{q-1}\left(1-\frac{q}{q^2-j}\right)\right)
       =(1-e^{-1}+o(1))q .
\]
The log of the product is $-1+O(1/q)$ by summing the first-order expansion. The excess leakage in bits is $\tfrac12\log_2 n+\log_2(1-e^{-1})+o(1)$. Both systems have the same recovery requirement, private capacity and final public context.
\subsection{Two audit languages with identical state counts}
Encode the legal histories as the two-letter language $\{x\,e:x\in e\}$ over disjoint fact and context alphabets. Compare a six-cycle with two disjoint triangles, taking vertices as facts and edges as contexts. Both complete minimal deterministic automata have nine states. Indeed, each has six distinct nonempty single-prefix residuals (the incident edge letters), in addition to the initial residual, the accepting residual $\{\varepsilon\}$, and the empty residual. Both have six facts, six contexts and degree two at every fact.
\par At $k=3$ the triangles have factor one by Corollary~1. In the six-cycle, a nonempty proper $s$-set hits at least $s+1$ edges, with equality for a contiguous segment. Equation~(\ref{eq:exp}) gives $4/3$. The distinction survives equal alphabet sizes, equal degrees, and equal minimal state counts. The missing invariant is the incidence of each feasible receipt support with the future contexts.
\section{What changes when records share private storage?}
At a joint checkpoint, two facts are available before either context is known. Each fact takes three values. Each context later reveals a pair containing its fact. The joint fact space is the $3\times3$ grid. The nine possible joint contexts are its $2\times2$ coordinate rectangles. Every receipt must permit exact reconstruction of the pair.
\par Compare two memory architectures with four private configurations. In the first, each coordinate must be recoverable from the public receipt and its own private bit. A receipt's support must have at most two values in each coordinate, hence must fit inside a $2\times2$ rectangle. This covers correlated encoders as well as independent ones. In the second, the reviewer can use both bits together to recover the pair. Any support of size at most four is then feasible.
\par\lead{Theorem 3} The optimum late-context guessing factors are $9/4$ with coordinate recovery and $2$ with pooled recovery. Both early-context optima equal one.
\begin{center}
\begin{tabular}{@{}lrrrr@{}}\toprule
Support size $s$ & 1 & 2 & 3 & 4\\\midrule
Fewest contexts hit, coordinate recovery &4&6&8&9\\
Fewest contexts hit, pooled recovery &4&6&6&8\\\bottomrule
\end{tabular}
\end{center}
\lead{Proof} Row and column permutations act transitively on facts and on contexts, with $n=m=9$. Apply (\ref{eq:exp}) to each admissible support family. A single cell meets four contexts. Two aligned cells meet six, and two unaligned cells meet seven. Three cells in a line meet six, whereas three in a $2\times2$ box occupy an L-shape and miss only the rectangle whose omitted row and column cover the L. A four-cell support cannot miss two distinct rectangles: their union has size at least six, leaving at most three cells. Four cells outside one rectangle can attain eight hits. A four-cell support inside a $2\times2$ box fills the box and meets all nine rectangles. Minimizing the hit counts divided by $s$ gives the stated factors. Early disclosure supplies the candidate rectangle, so one bit per coordinate suffices.\hfill$\square$
\par \begin{center}
\begin{tabular}{cc}
$\begin{matrix}\blacksquare&\blacksquare&\square\\\blacksquare&\blacksquare&\square\\\square&\square&\square\end{matrix}$\hspace{15mm} &
$\begin{matrix}\blacksquare&\blacksquare&\blacksquare\\\square&\square&\square\\\square&\square&\square\end{matrix}$\\[2pt]
{\small Four candidates, nine contexts hit}\hspace{9mm} & {\small Three candidates, six contexts hit}
\end{tabular}
\end{center}
The pooled optimum can also be attained with a full row as the candidate set, using three of the four private configurations. Choosing the support shape is part of the optimization. A coordinate-by-coordinate audit design can incur excess leakage even when both architectures receive exactly two protected bits.
\subsection{The repeated-record experiment}
At one joint checkpoint the encoder sees all $t$ facts. All $t$ context labels arrive only after the block receipt has been fixed. The observer learns the full label tuple. A pool with $K$ distinguishable private states must support exact reconstruction of the whole fact tuple. The objective is the worst-context maximal leakage of the block, divided by $t$ to report bits per record. This timing allows the encoder to allocate private capacity across coordinates.
\section{The exact rate for three facts}
Write $C_t$ for the minimum late factor at $K=2^t$. Index each two-element context by the symbol it omits. A vector of context labels is consistent with a fact vector precisely when the two vectors differ in every coordinate. Thus the public contexts hit by a support $A\subseteq[3]^t$ form its vertex neighborhood $N(A)$ in the direct graph power $K_3^{\times t}$. Separate transitivity on facts and contexts still holds. Equation~(\ref{eq:exp}) gives
\[
 C_t=\min_{1\le |A|\le 2^t}\frac{|N(A)|}{|A|}.
\]
\lead{Theorem 4} Let $r=1-\log_3 2$. For every positive integer $t$,
\[
 2^{tr}\le C_t\le 2^{\lceil tr\rceil}.
 \qquad\text{Consequently}\qquad
 \lim_{t\to\infty}\frac{\log_2 C_t}{t}=1-\log_3 2=0.3690702464\ldots .
\]
\lead{Proof} Write $\mu(A)=|A|/3^t$. The vertex-isoperimetric inequality for powers of cliques [11,12] gives $\mu(N(A))\ge \mu(A)^{\log_3(3/2)}$. At $|A|\le2^t$, dividing by $\mu(A)$ gives $|N(A)|/|A|\ge(3/2)^{t\log_3 2}=2^{tr}$. For the upper bound, put $j=\lfloor t\log_3 2\rfloor$. Keep the first $j$ ternary symbols in private memory and put the other $t-j$ symbols in the public receipt. The private pool uses $3^j\le2^t$ states. For each public coordinate the later context leaves exactly two possibilities, so the conditional guessing factor is $2^{t-j}$. Rounding costs less than one bit per block.\hfill$\square$
\subsection{Exact finite optima}
Let $B_t(s)$ be the minimum number of contexts hit by an $s$-element support. Taking a subset can only shrink a neighborhood, so the minimum for size at least $s$ equals that for size exactly $s$. The finite profile recursion of Alon and Defant [12, Thm.~1.4] specializes, with $M=3^{t-1}$ and $B_0(0)=0$, $B_0(1)=1$, to
\[
 B_t(s)=\begin{cases}
 2B_{t-1}(s),&0\le s\le M,\\
 2M+B_{t-1}(\lceil (s-M)/2\rceil),&M<s\le3M .
 \end{cases}
 \quad C_t=\min_{1\le s\le2^t}B_t(s)/s .
\]
\begin{center}\small
\begin{tabular}{@{}lrrrrrr@{}}\toprule
Block length $t$&1&2&3&4&6&8\\\midrule
Optimal factor&$3/2$&$2$&$9/4$&$16/5$&$5$&$8$\\
Bits per record&0.5850&0.5000&0.3900&0.4195&0.3870&0.3750\\\bottomrule
\end{tabular}
\end{center}
A longer block need not improve its immediate predecessor per record. The three-to-four reversal illustrates the effect of an integer private-state budget. We compared the recurrence at every admissible support size with direct enumeration for $t=1,2,3$, including all 3,505,698 nonempty eligible supports in the third power.
\section{An audit workload derived from OpenSSL}
An archive records which cipher suite an operation used. The trusted auditor must recover the exact suite. Later the archive reports one of several applicable policy labels. The observer has access to the archive and the label. Network peers, wire observers with the negotiated suite, and observers of the private index are outside this experiment.
\par We queried OpenSSL 3.5.7 locally using \texttt{openssl ciphers -s -tls1\_2} [9]. The workload selects ECDHE suites with AES-GCM or ChaCha20 and groups them by authentication family or bulk cipher. This yields six catalogue entries and five selected policy labels. The labels and the time at which they are revealed are experimental choices. They are not claims about TLS negotiation or policy frequencies in deployments.
\begin{center}\small
\begin{tabular}{@{}lccc@{}}\toprule
& AES-128-GCM & AES-256-GCM & ChaCha20-Poly1305 \\\midrule
ECDHE with ECDSA authentication & one suite & one suite & one suite \\
ECDHE with RSA authentication & one suite & one suite & one suite \\\bottomrule
\end{tabular}
\end{center}
Each row and each column is a possible future context. We minimized the worst-context score using a channel-form linear program independent of the cover derivation. We enumerate all size-$k$ public lists and allow arbitrary row-specific probabilities on each. Any smaller support can be padded to a fixed size-$k$ list, preserving decodability; merging receipts bearing the same padded list can only reduce the score. An epigraph variable for each context/list pair bounds its largest row probability. HiGHS found numerical solutions. We reconstructed rational primal and dual vectors, verified every inequality and row sum, and checked equality of their objectives. All six optima below therefore have exact finite certificates.
\begin{center}
\begin{tabular}{@{}rccc@{}}\toprule
Private states $k$ & Late factor & Early factor & Excess bits $\log_2(\caus/\know)$\\\midrule
1 & 3 & 3 & 0.0000\\
2 & $9/5$ & $3/2$ & 0.2630\\
3 & $8/5$ & 1 & 0.6781\\
4 & $4/3$ & 1 & 0.4150\\
5 & $6/5$ & 1 & 0.2630\\
6 & 1 & 1 & 0.0000\\\bottomrule
\end{tabular}
\end{center}
\lead{A hand certificate at $k=3$} By symmetry, an encoder may choose the entire row containing $x$ with probability $2/5$, and an L-shaped triple containing $x$ uniformly with probability $3/5$. The row-only mechanism has row/column factors $(1,2)$, and the uniform L mechanism has $(2,4/3)$ by incidence counting. The mixture gives $8/5$ on both. Conversely, every $s$-cell support for $1\le s\le3$ intersects at least $4s/3$ of the five row and column labels. Average the objective of (\ref{eq:lp}) over those labels and use $\sum_A |A|w_A=6$ to obtain the matching lower bound $(4/15)6=8/5$.
\par\lead{Controls} One private state forces public recovery of the suite. Six states allow a constant receipt. The timing penalty is largest at three states among these six choices. This is the smallest capacity at which an early recorder can hide the fact inside any one of the five contexts. Across all complete $d$-uniform families on $2\le n\le6$ we checked all 90 $(n,d,k)$ choices against (\ref{eq:complete}). All 164 budget choices over the 41 labeled four-vertex simple graphs without isolated vertices agreed with Corollary~1. We also checked both nine-state automata and enumerated all grid supports of sizes one through four. The source bundle contains frozen catalogue data, a live catalogue option and exact certificate checks. These measurements concern attainable audit-channel leakage. They do not estimate traffic prevalence or end-user harm.
\section{A rate formula for symmetric fact spaces}
Now let the finite incidence structure have $n$ facts. Assume a finite permutation group $G$ preserves the incidence relation and acts transitively on facts. Contexts may split into different orbits $E_1,\ldots,E_h$. For a nonempty fact support $A$ define
\[
 f_i(A)=\frac{n}{|A|}\frac{|\Gamma(A)\cap E_i|}{|E_i|},
 \qquad f_q(A)=\sum_{i=1}^h q_i f_i(A), \qquad q\in\Delta_h .
\]
Each $f_i(A)$ is the factor on type $i$ when a uniformly translated copy of $A$ containing the true fact is the public receipt. Transitivity normalizes each row of this channel. Every type covers the facts uniformly, so $f_i(A)\ge1$. Let $C_t(b)$ denote the optimal factor for $t$ facts with a pool of $\lfloor2^{tb}\rfloor$ states. All logarithms in this section are base two.
\par\lead{Theorem 5} For $0\le b\le\log n$ the exact asymptotic rate is
\begin{equation}
 \lim_{t\to\infty}t^{-1}\log C_t(b)
 =\max_{q\in\Delta_h}\ 
   \min_{\substack{\theta\in\Delta(2^V\setminus\{\varnothing\})\\
          \sum_A\theta_A\log|A|\le b}}
   \sum_A\theta_A\log f_q(A).\label{eq:orbitrate}
\end{equation}
When there is just one context orbit, the right side is the lower convex envelope of the one-record points $(\log |A|,\log f_1(A))$, optimized over mean private cost at most $b$. Multiple context orbits require the common adversarial mixture $q$. This is a proof device for the worst label tuple; the security experiment assumes no label frequencies.
\subsection{Converse for arbitrary pooled encoders}
Lift $q$ to a probability $Q$ on labels by distributing its mass uniformly within each orbit. For $0<p\le1$ let $d_p=\min_{A\ne\varnothing} Q(\Gamma(A))/|A|^p$. For any nonnegative vector $u$ indexed by facts, integration over the level sets of $u$ and the triangle inequality in $\ell_{1/p}$ give
\[
 \mathbb E_{e\sim Q}\max_{x\in e}u_x
 \ \ge\ d_p\int_0^\infty |\{x:u_x>z\}|^p\,dz
 \ \ge\ d_p\Big(\sum_x u_x^{1/p}\Big)^p.
\]
For a block support $D$ let $D_x$ be its last $t-1$ coordinates after fixing the first to $x$. At first label $e$ the neighborhood contains the union of the tail neighborhoods for $x\in e$. Bounding the probability of that union below by its largest member and applying the display to $u_x=|D_x|^p$ proves by induction that $Q^t(\Gamma(D))\ge d_p^t|D|^p$.
\par Average the block cover (\ref{eq:lp}) over $Q^t$ and use $\sum_D |D|w_D=n^t$. Every $D$ has size at most $2^{tb}$. Setting $\lambda=1-p$ yields the lower rate bound
\[
 \min_{A\ne\varnothing}\big[\log f_q(A)+\lambda\log|A|\big]-\lambda b .
\]
Maximizing over $\lambda\ge0$ is finite linear-program duality for the inner minimum in (\ref{eq:orbitrate}). Values $\lambda>1$ cannot improve on $1$: a singleton minimizes $n Q(\Gamma(A))|A|^{\lambda-1}$ there and $b\ge0$. The endpoint $p=0$ follows by continuity. The bound holds at every block length for every $q$.
\subsection{Achievability and the timing of labels}
Fix a distribution $\theta$ on support shapes with mean cost at most $b$. Allocate $t\theta_A$ records to each shape. Round the counts of nonsingletons down and fill the spare coordinates with a singleton; this preserves the budget and changes the empirical weights by $O(1/t)$ for a fixed finite incidence structure. Choose a uniform random permutation of that allocation, independent of the facts. Include the permutation in the public receipt. In each coordinate emit a uniform group translate of its assigned support containing the fact; store the indices jointly. The product of the support sizes is at most $\lfloor2^{tb}\rfloor$ and suffices for exact decoding.
\par To bound the worst label tuple, first choose its types independently with distribution $q$. Conditional on the public allocation, the factors multiply. Averaging over the types produces exactly $\prod_A f_q(A)^{t\theta_A}$ for integral counts. The public permutation makes the factor of a fixed type string depend only on its empirical distribution. Under independent sampling with that empirical distribution, its type class has probability at least $(t+1)^{-h}$: the empirical counts form a multinomial mode, and there are at most $(t+1)^h$ count vectors. Thus the maximum factor over type strings is bounded above by
\[
 (t+1)^h\max_q\prod_A f_q(A)^{t\theta_A}.
\]
After taking logarithms and dividing by $t$, rounding and the polynomial prefactor vanish. The objective is linear in $\theta$ and concave continuous in $q$, on compact convex sets. Sion's minimax theorem [13] exchanges the two optimizations and matches the converse.\hfill$\square$
\subsection{Why the permutation is public}
A fixed assignment can align the most revealing support type with the worst future label at every position. Permuting the assignment makes tuples with the same type counts exchangeable. Since the permutation is part of the receipt, its description costs no private states. The exogenous-label assumption still matters: a label chooser who can react to the observed permutation can restore the adverse alignment. Equation~(\ref{eq:orbitrate}) applies when labels are selected independently of the receipt coins conditional on the facts.
\subsection{A check using one context orbit}
For the three-fact system the one-record cost/leakage points are $(0,1)$, $(1,\log_2(3/2))$ and $(\log_2 3,0)$. The middle point lies strictly above the segment joining the other two. At a budget of one bit per record, the lower convex envelope has height $1-\log_3 2$. This recovers Theorem~4 by a second proof. The capacity is allocated by storing whole ternary facts in some positions and publishing whole facts in the rest. A scheme that encodes each fact with its own private bit pays the strictly larger one-record rate.
\section{Pooled leakage in the OpenSSL policy grid}
Return to the six-suite workload with a pool of $3^t$ states. Let $q$ be the test distribution's total mass on row labels. A support of size $s$ meeting $r$ rows and $c$ columns has
\[
 f_q(A)=\big[3qr+2(1-q)c\big]/s.
\]
\lead{Corollary 2} Let $q_*$ be the unique solution in $(0,1)$ of
\[
 2-q_*/2=(2+q_*)^{1-\log_6 2}.
\]
Then the optimal late leakage in bits per record tends to
\[
 R_{\rm grid}=(1-\log_6 3)\log_2(2+q_*)=0.5104636788\ldots .
\]
Knowing all labels early gives zero additional leakage, since each coordinate has at most three candidates.
\par\lead{Proof} Three shapes suffice for a matching construction: a singleton, a two-cell horizontal pair, and the full grid. Their respective factors are $2+q$, $2-q/2$, and $1$. The two sides of the defining scalar equation vary strictly in opposite directions, and their ordering reverses between zero and one. At its root all three points $(\log_2 s,\log_2 f_q(A))$ lie on the line from $(0,\log_2(2+q_*))$ to $(\log_2 6,0)$.
\par For a lower bound, every other support lies above this line. The inequalities to verify are $f_{q_*}(A)\ge(2+q_*)^{1-\log_6 s}$. At $s=2$ the horizontal pair minimizes the factor for $q_*>2/5$. At sizes $3,4,5$ the minima near $q_*$ are $2-q_*$, $1+q_*/2$ and $6/5$ respectively, by counting occupied rows and columns. Outward-rounded interval arithmetic on $0.49584<q_*<0.49586$ gives strict margins greater than $0.079$, $0.017$ and $0.102$. Sizes one and six are identities. The line therefore supports the inner optimization of (\ref{eq:orbitrate}) from below at $b=\log_2 3$.
\par For achievability, use proportions $u,v,w$ of the three shapes above. Set $v/u=(4-q_*)/(2+q_*)$, $u+v+w=1$ and $v+w\log_2 6=\log_2 3$. These give positive fractions approximately $(0.207890,0.291876,0.500234)$. The derivative of $u\log_2(2+q)+v\log_2(2-q/2)$ vanishes at $q_*$. Concavity makes it the maximum over $q$. Its value attains the supporting line at the required budget, proving equality.\hfill$\square$
\subsection{Finite-block check independent of the rate proof}
For two records, the exact optimum is $24/11$, or $0.562765441$ bits per record. To check it, observe that a tuple of labels projects a support onto one selected axis per record. Compressing a support down any coordinate preserves its size and can only shrink every such projection: a union of compressed fibers has the size of the largest fiber. Repeated compression terminates at a coordinatewise downset. Averaging each support over fact-value and record permutations preserves coverage and reduces the worst score. Conversely, each orbit-averaged shape is an attainable encoder; any symmetrized cover is a mixture of these encoders (a shape's mixing weight is its total cover weight times its size divided by $6^t$). At $t=2$ this leaves 438 nonempty downsets of size at most nine and 127 distinct cost vectors across the three label-count types. Exact rational primal and dual vectors certify the optimum. The one-record factor $8/5$, the two-record certificate, and the limiting saddle value use separate calculations.

\section{A working receipt encoder}\label{sec:codec}

\paragraph{Interface and accounting.} Setup fixes an ordered event alphabet, a private capacity $k$, and a rational cover from (\ref{eq:lp}). The recorder runs $\mathsf{Seal}(x)$ once per event or joint checkpoint. It sends its first output to a public archive and its second output to a separate confidential store. The reviewer passes those two outputs to $\mathsf{Recover}$. An application must associate the two sides with the same event. We expose the archive to a passive reader; the association, transport and storage are trusted. The scope of a receipt is the pre-agreed catalogue and block length. The private index counts against the budget even if an implementation stores it in a byte with unused values.

\paragraph{Exact sampler.} Sort each support and the list of supports lexicographically. For fact $x$, take the least common multiple $D_x$ of the denominators of the positive $w_A$ for $A\ni x$. Lay consecutive intervals of lengths $D_xw_A$ over the integers $0,\ldots,D_x-1$. Draw one uniform integer, emit the selected support's identifier $i$, and retain the zero-based rank $j$ of $x$ in $A_i$. The decoder returns $A_i[j]$. Interval lengths sum to $D_x$, so the realized channel has $W(i\mid x)=w_{A_i}\mathbf{1}_{x\in A_i}$. Its score is exactly the fractional-cover objective including all observable fixed format bytes. This construction has zero decoding error and at most $k$ possible private ranks. The interval sampler avoids rounding probabilities in the encoder. It uses the operating-system random source through Python's \texttt{secrets.randbelow} [16].

\paragraph{Wire format.} The finite encoder emits a three-byte version prefix, a 16-byte prefix of the SHA-256 digest of the public codebook serialization, and a big-endian fixed-width identifier. The confidential output holds a fixed-width, big-endian unsigned rank. The digest identifies the configuration. Authentication against alteration requires a separate integrity mechanism and threat model. In particular, a substituted valid rank or receipt can decode to a different fact. The proof counts public bytes read by the archive observer. It excludes implementation timing, the caller's plaintext buffers, and access to the private store. Distinct sampling of repeated copies of the same fact can expose more than a single receipt; retries should return the stored pair when the intention is one disclosure.

\subsection{A test vector and an optimal two-record encoder}
Index the first grid row by $0,1,2$ and the second by $3,4,5$, with columns in the order of the six-suite table. On input $4$ (ECDHE-RSA with AES-256-GCM), one possible public support is $(0,3,4)$ and the private byte is \texttt{02}. If the later label specifies RSA, the archive reader is left with $3$ and $4$. Reading private rank two recovers $4$ exactly. This receipt occurs with probability $1/10$ on each of its three possible inputs. The public suffix is \texttt{05}; the source bundle includes the complete hexadecimal test vector.

For a single event, the public codebook consists of two whole rows of weight $2/5$ each and twelve L-shaped triples of weight $1/10$ each. Each fact lies in one whole row and six L-shapes. A single uniform draw among ten tickets implements the optimum $8/5$ without an optimization solver at recording time.

For an ordered pair, write $H$ for a horizontal two-cell subset, $F$ for the four-cell strip formed by two full columns, and $G$ for the whole six-cell grid. First use the orbit of $H\times F$, including record swaps. Next use the orbit of
\[
 (\{0\}\times G)\;\cup\;(\{1\}\times\{0,1,2\}).
\]
The relabelings independently permute authentication rows and cipher columns in each record. For an orbit $\mathcal O$ of supports of size $s$, selected with probability $p$, its cover weight per member is $36p/(|\mathcal O|s)$. The table gives the factor of each orbit used alone. The mixed type averages over the two positional orders, which are exchanged by the construction.
\begin{center}\small
\begin{tabular}{@{}lrrrrrr@{}}\toprule
Orbit & Members & Size & Probability & Two columns & Mixed & Two rows\\\midrule
Horizontal pair and strip & 36 & 8 & $8/11$ & $2$ & $9/4$ & $9/4$\\
Nine-cell union & 48 & 9 & $3/11$ & $8/3$ & $2$ & $2$\\\bottomrule
\end{tabular}
\end{center}
All three conditional factors therefore equal $24/11$. The two-record lower certificate in the preceding section proves optimality. Each of the 36 inputs admits 44 equiprobable tickets. Exhaustive decoding of all 1,584 input/ticket pairs recovers the original ordered pair, and exact evaluation of all 25 ordered policy pairs gives the same factor. By comparison, encoding the two records separately gives $(8/5)^2=64/25$. Both architectures have nine joint private states.

\subsection{A pooled encoder for long blocks}
The block encoder uses three shapes: a singleton, a horizontal pair, and the full grid. At length $t$, round down the pair and full-grid counts from Corollary~2 and assign the remaining positions singletons. A final integer check enforces $2^v6^w\le3^t$, where $(u,v,w)$ are now \emph{integer counts}. Uniformly shuffle the multiset of shapes. At each pair position choose one of the two other entries in the event's row uniformly. The public message is a fixed prefix, the block length, and one six-bit support mask stored in a byte per position. The store holds a single mixed-radix index. Starting at $J=0$, update $J\gets J|A_i|+j_i$. Repeated division in reverse order recovers every $j_i$. The receipt masks disclose the shuffled schedule, as permitted in the proof.

For $j$ future row labels and $t-j$ column labels the \emph{implemented} factor has an exact integer-coefficient calculation:
\[
 L_t(j)=\frac{[z^j](2+3z)^u(4+3z)^v(1+z)^w}{2^v\binom tj}.
\]
Indeed each singleton contributes factors $(2,3)$ for column and row, each horizontal pair $(2,3/2)$, and each full grid $(1,1)$. The public shuffle averages their products over the assignments of the $j$ row positions. The maximum over $j$ is the attained factor of this schedule. This calculation also checks rounding losses at finite block lengths.
\begin{center}\small
\begin{tabular}{@{}rrrrrr@{}}\toprule
Records & Singletons & Pairs & Full grids & Worst row count & Bits/record\\\midrule
8&2&2&4&7&0.542481\\
32&7&9&16&20&0.517191\\
128&27&37&64&68&0.512342\\
512&107&149&256&260&0.511238\\
1,024&214&298&512&520&0.511228\\\bottomrule
\end{tabular}
\end{center}
Enumeration of all inputs, distinct schedules, pair choices and subsequent labels for $t\le4$ reproduced the coefficients from actual receipt bytes. For larger $t$ the table uses the exact polynomial and an end-to-end encoder/decoder check. The schedule attains the rate $0.5104636788\ldots$ asymptotically. The finite optimal one- and two-record encoders above have separate entry points.

\section{A 65-suite catalogue and the cost of protecting individual policies}\label{sec:large}
\paragraph{Data and protocol.} We enumerated every suite returned by \texttt{openssl ciphers -s -tls1\_2 -stdname -v ALL} on the recorded build [9]. The 65 suites can be used with TLS 1.2 under the catalogue command; some names originated in earlier protocol versions. Each event has four labelled attributes: authentication, key exchange, bulk cipher as rendered by OpenSSL, and MAC category. Distinct values yield 21 labelled contexts. We added two externally specified lists, the four suites recommended in RFC~9325 \S4.2 and the RFC~7905 \S3 assignments present locally [14,15]. The workload contains 23 labelled contexts and 267 event/label incidences. RFC~7905's three available entries also form the existing ChaCha20 bulk-cipher set; we preserve both policy names and count 22 distinct subsets. Each suite counts once in the domain. No network trace or policy-frequency sampling enters these computations.

\paragraph{Codebook synthesis.} For each of the 24 orders of the four attributes, form groups using the first zero, one, or two attributes. Within each group sort by the complete attribute order, breaking ties by IANA name. On a cycle of length $g$ use every window of length $h=\min(k,g)$, each with weight $1/h$; merge repeated windows by adding weights. Each fact belongs to $h$ windows counting multiplicity. Mix these 72 candidates using a linear program on the 21 attribute labels. Recover rational mixing coefficients, then compute the achieved factors exactly. The two standards-labelled scores below are evaluated after fitting. One of those sets already has an equivalent attribute label, so this is a label holdout with only one new subset. For a universal lower bound use the 40-suite RSA authentication context. Every exact-recovery encoder has $\rho\ge40/k$ for the tested capacities.
\begin{center}\small
\begin{tabular}{@{}rrrrrr@{}}\toprule
$k$&Receipts&Attained factor&Lower bound&RFC 7905&RFC 9325\\\midrule
1&65&40&40&3&4\\
2&65&20&20&3&3\\
3&65&$40/3$&$40/3$&3&$8/3$\\
4&116&10&10&3&$5/2$\\
6&191&6.9744&6.6667&2.8462&2.3590\\
8&228&5.7777&5&2.5450&2.0705\\
12&186&4.1426&3.3333&2.3295&2.0092\\
16&191&3.4414&2.5&2.1547&1.7405\\\bottomrule
\end{tabular}
\end{center}
The first four rows reach the unrestricted optimum. The remaining rows give attainable upper bounds with explicit gaps; the source bundle includes the exact fractions and all eight public codebooks. The stock $k=3$ encoder saturates the global optimum while every observed receipt fully distinguishes the three candidates once the RFC~7905 label is known. A targeted redesign removes part of this disclosure without increasing the global score.

\subsection{A stability inequality and a simultaneous optimum}
\paragraph{Proposition 6.} Write $L_e$ for the factor on context $e$. For any two contexts $D,e$ with $e\cap D\ne\varnothing$ and $e\setminus D\ne\varnothing$,
\begin{equation}
 L_e\ \ge\ 2-k\big(L_D-|D|/k\big).\label{eq:slack}
\end{equation}
For a fractional cover the difference $L_D-|D|/k$ equals
$\sum_{A\cap D\ne\varnothing} w_A(1-|A\cap D|/k)$. Every support meeting both sides of $D$ contributes at least $w_A/k$. Choose $x\in e\cap D$ and $y\in e\setminus D$. The cover weight of supports containing either point is $2-\sum_{A\supseteq\{x,y\}} w_A$, by row normalization. The common supports cross the cut. Bounding their total weight by the previous sum proves (\ref{eq:slack}). The level-set reduction of Theorem~1 preserves every contextual score simultaneously, so the inequality holds for arbitrary randomized encoders.\hfill$\square$

\paragraph{Targeted design.} At $k=3$, both RFC sets cross the RSA cut. Consequently $L_e\ge2$ on both whenever the global optimum $40/3$ is attained. Pin three disjoint triples with unit weight. The first groups the two RSA members of RFC~9325 with the ECDHE-RSA AES-128-CBC-SHA suite. The second groups the two RSA members of RFC~7905 with the DHE-RSA AES-128-CBC-SHA suite. The third groups the three non-RSA members of the union of both standards lists. Refit the residual cover on the remaining 56 suites. The resulting 59-receipt codebook attains
\[
 (\rho,L_{\rm RFC\,7905},L_{\rm RFC\,9325})=(40/3,\,2,\,2).
\]
All three numbers have matching lower bounds under the global-optimality constraint. In particular, requiring a factor of at most $3/2$ on either cross-cut profile would force $\rho\ge 40/3+1/6$. The original held-out results above and the standards-aware refit are separate experiments. Exact inspection of the refitted encoder's 177 nonzero event/receipt memberships confirms normalization, private recovery and all 23 contextual factors. The guarantee concerns a reader of the audit archive; the TLS catalogue supplies the domain, and these measurements do not test TLS confidentiality.

\section{When the reader chooses what the audit discloses}\label{sec:feedback}
Consider a policy engine that answers authenticated questions about a logged event. An archive reader can inspect its permanent receipt, select one attribute, see its value, and request a second attribute on selected outcomes. The engine charges for the second request. Two systems with the same marginal frequency of follow-ups can have different disclosure probabilities because the reader can spend those requests on ambiguous answers. We derive the best receipt against this decision rule, then remove the recorder's knowledge of the reader's budget.

\paragraph{Experiment.} The finite event set $X$ has a public prior $\pi$ of full support. Two public maps $f_0,f_1$ specify truthful attribute answers. The recorder publishes $r$ using $W(r\mid x)$ and retains a private role in $[3]$ sufficient for zero-error recovery. After seeing $r$, the reader chooses $i$, observes $f_i(x)$, then either guesses $x$ or obtains $f_{1-i}(x)$ before guessing. A behavioral policy $\sigma$ can randomize at both stages. Let $C\in\{0,1\}$ indicate the second request, whose average budget $b$ is public unless stated otherwise. Set
\[
 V_b^\pi(W)=\max_{\sigma:\,\mathbb E C\le b}\Pr_{\pi,W,\sigma}[\widehat X=X],
 \qquad M_b(\pi)=\inf_{W:\,|\operatorname{supp} W(r\mid\cdot)|\le3\ \forall r}V_b^\pi(W).
\]
The observer can spend the budget selectively across receipts. The experiment scores absolute identification probability for the given prior. Equation~(\ref{eq:metric}) instead measures the amplification after an exogenous disclosure. Their numbers have different units. The free and paid answers, and the identity of the branch on which a request was made, belong to the decision history. The protected role never enters that history.

\paragraph{A counterexample with no follow-up.} On a uniform $3\times3$ grid let the receipt reveal the entire row containing the event with probability $1/2$ and the entire column otherwise. The private rank has three values. A reader committing in advance to either coordinate identifies the cell with probability $(1+1/3)/2=2/3$. A reader who sees the receipt first can always request the other coordinate and identify the cell. Here $b=0$. Changing who chooses the question already changes the inference problem before any second query is issued.

\subsection{From question partitions to privacy curves}
Boreale and Pampaloni [17, Sect.~6.1] evaluate a finite adaptive strategy by recursing over observed answers. Kamenica and Gentzkow [18, Prop.~1] characterize implementable posterior laws by their barycenter. The following reduction combines these tools with a hard three-state recovery limit and an expected query cost. For a posterior $p$, write $p(D)=\sum_{x\in D}p(x)$ and $m_p(D)=\max_{x\in D}p(x)$ with empty maximum zero. Let $\mathcal P_i$ be the partition into $f_i$-fibres. At second-query price $\lambda$ the Bellman value of a public posterior is
\begin{equation}
 B_\lambda(p)=\max_{i=0,1}\sum_{D\in\mathcal P_i}
 \max\left\{m_p(D),\
 \sum_{E\in\mathcal P_{1-i}} m_p(D\cap E)-\lambda p(D)\right\}.
 \label{eq:priced-audit}
\end{equation}
The inner maximum occurs after the first answer. The price multiplies the probability of reaching that branch. Moving either operation outside its sum prices a different policy. For a function $B$ let $\operatorname{co}_3 B(\pi)$ be the least $\sum_j t_j B(p_j)$ over posterior mixtures with $\sum_j t_jp_j=\pi$, $\sum_jt_j=1$ and $|\operatorname{supp}p_j|\le3$. Then
\begin{equation}
 M_b(\pi)=\min_{0\le\lambda\le1}\{\lambda b+\operatorname{co}_3 B_\lambda(\pi)\}.
 \label{eq:price-envelope}
\end{equation}
Indeed, mixtures of deterministic decision trees form a compact polygon of expected cost and success for a fixed channel. Linear programming duality for its single budget constraint gives a price and the backward recursion (\ref{eq:priced-audit}). A price above one cannot improve the bound because success can increase by at most the probability that the reader pays. Given a posterior mixture, the recorder implements $W(j\mid x)=t_jp_j(x)/\pi(x)$ and privately retains the rank of $x$. Conversely each receipt induces this mixture by Bayes' rule. Minimizing over both the price and the mixture gives (\ref{eq:price-envelope}). The feasible posterior graphs at fixed price are compact. Carath\'eodory's theorem bounds the needed mixture size by $|X|+1$ using the graph coordinate for the value. Joint continuity and compactness then give attainment, including at the two budget endpoints. This reduction gives a design criterion for arbitrary finite attribute maps. The next results solve it on a specific geometry.

\paragraph{Generating the equations from the permitted operations.}
The recorder fixes a support, the reader may stop and guess, a question partitions that support, and a second question incurs one unit of cost. Those four facts suffice to generate the price equations below. Let $A$ be a set of possible events and let a deterministic transcript tree have two vectors indexed by $A$: $a_x$ indicates a correct final guess on $x$ and $c_x$ counts its paid questions on $x$. For a sequence of available stage prices $d$, define the tree profiles $\mathcal T_d(A)$ by
\begin{align}
 \mathcal T_{\varnothing}(A)&=\{(\mathbf e_x,0):x\in A\},\nonumber\\
 \mathcal T_{(q,d)}(A)&=\mathcal T_{\varnothing}(A)\ \cup\
 \bigcup_f\left\{\left(\sum_D a_D,\ q\mathbf1_A+\sum_D c_D\right):
 (a_D,c_D)\in\mathcal T_d(D)\ \text{for each }D\right\}. \label{eq:treegen}
\end{align}
Here $D$ ranges over the nonempty answer cells of $f$ in $A$ and child vectors are extended by zero. Taking $d=(0,1)$ at the root gives our two-stage game. Every legal policy first guesses or chooses a test and then chooses a continuation separately on each answer; induction gives both directions of (\ref{eq:treegen}). Randomized policies convexify the profiles. Finite-horizon value vectors have a classical precedent [25]. Here each vector retains the eventwise expenditure as well as correctness, because the recorder subsequently chooses which posterior the reader receives.

Let $P_A(\lambda)$ be the convex hull of the vectors $a-\lambda c$ from this construction. Its support function $h_A(v)=\max_{z\in P_A(\lambda)}\langle v,z\rangle$ evaluates an unnormalized posterior $v$. This explains exactly when the older fractional-cover reduction survives a change of protocol. If $v=\sum_j v_j$ and publication also identifies $j$, the priced gain from this refinement is
\begin{equation}
 \sum_j h_A(v_j)-h_A\!\left(\sum_jv_j\right)\ \ge0,
 \qquad
 \text{equality}\ \Longleftrightarrow\
 \bigcap_j\operatorname*{arg\,max}_{z\in P_A(\lambda)}
       \langle v_j,z\rangle\ne\varnothing .\label{eq:shared-tree}
\end{equation}
Indeed, an optimizer for the sum attains equality only if it optimizes every summand. In the fixed-context game the single-guess functional is $\max_{x\in e}v_x$. The point attaining the largest coordinate on $e$ attains the maximum on every nested level indicator of $v$. This is the structural reason the cover in Theorem~1 is exact. In a three-cell corner with posterior $(1/2,1/4,1/4)$ and $\lambda=1/3$, stopping after the first answer and paying on its ambiguous branch both give $3/4$. Refining the column into a uniform triple of mass $3/4$ and the pivot of mass $1/4$ raises the value to $(3/4)(7/9)+1/4=5/6$. The old representation loses exactly $1/12$ at this price. The failure can be read off from the absence of a common optimizing tree.

\paragraph{From a finite menu to an exact curve.}
Suppose a group acts transitively on events and permutes the permitted questions up to answer relabeling. For the uniform prior, let $A$ range over $k$-point supports, allowing a posterior to assign zero to some members. Define the local matrix $D_A(\lambda)$ with rows $a-\lambda c$ from (\ref{eq:treegen}). Then
\begin{align}
 \gamma_A(\lambda)
 &=\min_{p\in\Delta(A)}\max_T (D_Ap)_T
   =\max_{\sigma\in\Delta(\mathcal T)}\min_{x\in A}(D_A^\top\sigma)_x,\nonumber\\
 G(\lambda)&=\min_A\gamma_A(\lambda),
 &M_b&=\min_{0\le\lambda\le1}\{G(\lambda)+b\lambda\}. \label{eq:generating-price}
\end{align}
The middle game equality is finite matrix duality. The lower bound for $G$ holds receipt by receipt. Conversely, translating any minimizing posterior over the group gives a mixture with uniform barycenter without changing its priced value. Bayes inversion implements this mixture, proving the upper bound and the last formula by the price duality above. This step eliminates the unknown encoder before any candidate codebook is proposed. A nonuniform population retains the barycenter constraint of (\ref{eq:price-envelope}).

The calculation uses only the partitions induced on each $A$. Isomorphic restrictions have identical local games, and averaging $p$ over a restriction's automorphisms preserves feasibility and can only reduce its worst score. On a fixed optimal matrix-game basis the equilibrium constraints are a linear system with entries affine in $\lambda$. Cramer's rule gives rational functions of $\lambda$; primal and dual signs specify their validity intervals. Taking the lower envelope in (\ref{eq:generating-price}) and comparing interval endpoints with the solutions of $G'(\lambda)=-b$ generates the budget curve. The code proposes bases numerically and certifies each by exact rational-function identities and polynomial root isolation. It verifies the whole price interval, including changes of basis. Continuity supplies the values at isolated endpoints.

For the eight-event, three-coordinate menu the compiler generates the following expressions from the same tree recursion. Here $p$ in the calculation above is a posterior and $\lambda$ denotes the scalar price.
\begin{center}\small
\begin{tabular}{@{}cll@{}}\toprule
Private roles & Generated $G(\lambda)$ & Generated interior transition\\\midrule
3 & $\max\{(1+\lambda)^{-1},\,2/3\}$ & $b=4/9$\\
4 & $\max\{3(1-\lambda)/4,\,1/2\}$ & $b=3/4$\\\bottomrule
\end{tabular}
\end{center}
The reciprocal in the first row has a specific origin. In a corner let $s$ be first-answer success when the pivot is heaviest and both arms have equal weight. The competing trees have values $s$ and $1-\lambda s$. Equalizing them gives $s=(1+\lambda)^{-1}$ until $s$ reaches $2/3$. Minimizing the generated branch plus $b\lambda$ gives $2\sqrt b-b$, with the change at $4/9$ coming from the derivative at $\lambda=1/2$. In the second row the minimizing posterior can stay uniform while the best tree switches. The two priced payoffs are affine, so the budget curve has a corner rather than a square-root branch. These facts explain why changing the decoder capacity changes even the algebraic form of the answer.

More generally the lower closed convex envelope of $G$ on $[0,1]$ is the part recoverable from all budget experiments:
\[
 (\operatorname{cl\,conv}G)(\lambda)=\sup_{0\le b\le1}\{M_b-b\lambda\}.
\]
To see this, apply the supporting-line representation of the lower convex envelope. Each tree has cost between zero and one, so $G$ is nonincreasing and $1$-Lipschitz, as is its lower convex envelope. Supporting slopes in $[-1,0]$ suffice, and (\ref{eq:generating-price}) gives their intercepts. Thus two such menus are indistinguishable by this entire family of optimal-budget measurements exactly when their envelopes coincide. Equality of a single endpoint is a weaker test.

\subsection{A sharp change in the optimal encoder}
Suppose $X$ is a full $m\times n$ rectangle, $m,n\ge2$, with uniform prior and the two coordinate queries. Write $M_b$ for its value. For $2/3\le s\le1$, set
\[
 h_b(s)=s+(1-s)\min\{1,b/s\}.
\]
\lead{Theorem 7} Each feasible channel has a single $s\in[2/3,1]$ such that $V_b^\pi(W)\ge h_b(s)$ for \emph{every} budget. For every such $s$ there is a channel attaining the entire curve $h_b(s)$. In particular,
\begin{equation}
 M_b=
 \begin{cases}
 2/3+b/2,&0\le b\le4/9,\\
 2\sqrt b-b,&4/9\le b\le1 .
 \end{cases}\qquad
 s_b=\max\{2/3,\sqrt b\}.
 \label{eq:sharp-feedback}
\end{equation}

\paragraph{Converse.} Condition on a receipt. A coordinate can distinguish the two most likely cells since their coordinates differ somewhere. Prefer an injective coordinate on the support whenever one exists. Otherwise the support consists of three corners of a rectangle. Among coordinates separating the two most likely cells choose one minimizing the mass of the ambiguous branch. If it leaves a pair, the remaining singleton is at least as likely as the less probable member of the pair. Writing $s_r$ for success after that first question and $u_r$ for the probability of an ambiguous answer gives $s_r\ge2/3$ and $u_r\le s_r$. Indeed, for a pair of masses $v\ge w$ and singleton $t\ge w$, one has $s_r=v+t$ and $u_r=v+w$. Supports of size at most two can be separated with one question. Let $s=\mathbb E s_r$ and $U=\mathbb E u_r$. The other coordinate resolves every remaining ambiguity. Requesting it on a fraction $\min(1,b/U)$ of ambiguous branches recovers that fraction of all remaining errors. If $U=0$ success is one. Otherwise $U\le s$ gives $V_b^\pi(W)\ge h_b(s)$. The first question used to define $s$ is independent of $b$.

\paragraph{Matching channel.} Choose three corners of a $2\times2$ subrectangle. Its pivot is the cell sharing a row with one arm and a column with the other. Give the event posterior probabilities
\[
 (p_{\rm pivot},p_{\rm horizontal},p_{\rm vertical})=(2s-1,1-s,1-s)
\]
and mix all oriented corners with equal receipt probabilities. Each cell occurs $(m-1)(n-1)$ times in each role, giving the uniform prior. Equivalently the recorder samples the private role with these three probabilities and samples one other row and column uniformly. It publishes the resulting oriented corner. Either first question gives one singleton and a pair, with ambiguous mass $s$, total error $1-s$, and gain $(1-s)/s$ per paid ambiguous branch. No selective policy can earn a higher marginal gain on any such branch. At $s=1$ the receipt determines the event and the same formula holds by continuity. This proves equality simultaneously for all $b$. On $s>b$, $h_b(s)=s+b/s-b$ has derivative $1-b/s^2$. On $s\le b$ it equals one. Minimization gives (\ref{eq:sharp-feedback}).\hfill$\square$

\paragraph{Why equal weights fail.} Under the uniform prior a flat receipt column gives a uniform posterior on at most three cells. The policy used in the converse leaves only equally likely pairs or singletons. Each paid ambiguous branch therefore corrects half a guess per unit of probability spent. The exact optimum over flat columns is $\min(1,2/3+b/2)$, attained by the uniform three-corner channel. It strictly exceeds $M_b$ for $4/9<b<1$. The level decomposition in Theorem~1 preserves the maxima on fixed contexts, but publishing its level gives the reader new information with which to allocate the follow-up budget. For the six-suite grid at $b=16/25$, the flat optimum is $74/75$. The posterior $(3/5,1/5,1/5)$ attains $24/25$. With this release the first question succeeds with probability $4/5$ and the second request earns $1/4$ extra correct guesses per paid ambiguous branch.

\paragraph{One mechanism when the budget is unknown.} If an encoder is fixed before the budget is specified, its additive regret at $b$ is $V_b^\pi(W)-M_b$. The best worst-budget regret is exactly
\begin{equation}
 \inf_W\sup_{0\le b\le1}(V_b^\pi(W)-M_b)=1/36.
 \label{eq:regret}
\end{equation}
For each $W$ the converse supplies the same $s$ at every budget. For a corner channel its maximum regret is $\max\{s-2/3,(1-\sqrt{s})^2\}$. To see this, on $[0,4/9]$ the difference is nonincreasing, on $[4/9,s]$ it is convex and takes its maximum at an endpoint, and on $[s,1]$ it is nonincreasing. The first term grows with $s$ and the second decreases. Equalizing them gives $s=25/36$. The encoder uses private-role probabilities $(14,11,11)/36$. The budgets $0$ and $25/36$ both witness the lower bound. This is one deployable channel for the entire interval.

\subsection{Which nonuniform event populations admit the same guarantee?}
Assume each pair of attributes identifies at most one event. The realized cells may form a proper subset of a rectangle. For a corner $A$ whose three cells are present, let $p_A(s)$ be its pivot-and-arms posterior, and let $K_s$ be the convex hull of all these vectors. The converse of Theorem~7 still applies. The uniform mixing argument need not.

\paragraph{Theorem 8 (equality criterion).} Fix $4/9<b<1$ and $s_*=\sqrt b$. Then
\begin{equation}
 M_b(\pi)=2\sqrt b-b\quad\Longleftrightarrow\quad \pi\in K_{s_*}.
 \label{eq:corner-prior}
\end{equation}
Any convex representation supplies a recorder by Bayes inversion. Equivalently, membership holds exactly when $\langle z,\pi\rangle\le\max_A\langle z,p_A(s_*)\rangle$ for every vector $z$ on the cells. If there are no corners the set is empty and one free question separates every support of size at most three after the receipt is seen.

\paragraph{Proof.} A mixture in $K_{s_*}$ has the same two possible branch types and the same marginal gain at every ambiguous answer, hence attains $h_b(s_*)$. Conversely, take an attaining channel from (\ref{eq:price-envelope}). Equality in the scalar minimization forces the unique $s=s_*$. As $0<b<s<1$, having $U<s$ would strictly improve the reader's uniform-thinning policy above $h_b(s)$. Thus $U=s$, and $u_r=s_r$ at every positive-probability receipt. An injective first question would give $u_r=0<s_r=1$. All remaining supports are corners. The arms must have equal probabilities and the pivot must be at least as likely as either arm. To verify this, if a maximal arm is heavier than the smallest atom, isolate that arm by its coordinate: this is a best first question and has $u_r<s_r$. If the pivot is maximal and the arms have different weights, isolating the heavier arm has the same strict effect. Ties among maximal atoms obey the same argument. Each remaining receipt therefore has the form $p_A(s_r)$. Its error per unit of ambiguous mass is $(1-s_r)/s_r$. An interior fraction $b/U$ of every ambiguous branch was paid. If two of these rates differ, shifting a positive amount of spending from the smaller rate to the larger one gives a strict improvement at the same cost. All rates coincide, hence all $s_r=s_*$. The barycenter condition forces $\pi\in K_{s_*}$. The separating-vector statement is finite convex separation.\hfill$\square$

\paragraph{Two nonuniform tests on the six-suite grid.} Use coordinates $0,1$ for the rows and $0,1,2$ for the columns. Fix two disjoint corners with pivots $(0,0)$ and $(1,1)$ and arms $\{(1,0),(0,2)\}$ and $\{(0,1),(1,2)\}$ respectively. With $s=4/5$ and public mixture weights $1/3,2/3$, Bayes inversion attains $24/25$ at $b=16/25$ on the resulting nonuniform population. Now assign probability $7/20$ to each pivot and $3/40$ to each of the other four cells. Take $z$ to be the indicator of the two pivots. Their total prior mass is $7/10$. These cells differ in both coordinates, so within any one corner they can occur together only as its arms. Consequently $\max_A\langle z,p_A(4/5)\rangle=\max(3/5,2/5)=3/5$. This is a strict certificate of failure even though each individual cell passes the single-cell upper bound $3/5$. A feasible recalibration uses the same two public receipts with equal weights and $s=17/20$. Their barycenter is exactly the stated prior. The decoder stays exact and transcript evaluation gives
\[
            24/25 < M_{16/25}(\pi) \le h_{16/25}(17/20)=1637/1700 .
\]
The upper endpoint is attained by the recalibrated recorder. The exact optimum inside this interval is unresolved.

\paragraph{Sensitivity to a measured prior.} If $\|\pi-\pi'\|_{\mathrm{TV}}\le\delta$, each fixed behavioral policy has both its success and its expected cost changed by at most $\delta$ by contraction through the transcript channel. For $\delta\le b\le1$, a fixed encoder therefore obeys
\[
 V_{b-\delta}^{\pi}(W)-\delta\le V_b^{\pi'}(W)
       \le V_{\min(1,b+\delta)}^\pi(W)+\delta .
\]
Changing the budget in this bound is essential: a policy admissible under the nominal prior can exceed its average query quota after a prior shift. On the uniform six-suite grid shift mass $1/12$ from the bottom right to the top left cell, holding the $s=4/5$ encoder fixed. At $b=16/25$ the reader's success rises from $24/25$ to $1031/1050$. These are prescribed prior perturbations, not frequencies estimated from TLS traffic.

\subsection{What survives attribute collisions in the 65-suite catalogue}
In the catalogue, the attribute pair (authentication, bulk cipher) takes only 30 distinct values on 65 suites. Thus two answers need not identify the event. Let $\mathcal G$ be the joint partition. With uniform prior on $N$ events and arbitrary private capacity $k$, the exact full-budget endpoint is
\begin{equation}
 M_1(\pi)=\frac{1}{N}\sum_{G\in\mathcal G}\max(1,|G|/k).
 \label{eq:fibres}
\end{equation}
Indeed the reader can request both attributes. For each $G$, the mass-normalized contribution $\sum_r\max_{x\in G}W(r\mid x)$ is at least one by normalization and at least $|G|/k$ by the support bound. To attain both bounds, first compute the joint fibre containing the input, then uniformly release a subset of size $\min(k,|G|)$ of that fibre containing the input. The private rank decodes the event. Both answers are now fixed by the public subset. This design discloses the joint attribute intentionally. It optimizes event identification at full budget and would need a different objective if those attributes themselves were sensitive.

\begin{center}\small
\begin{tabular}{@{}lrrrr@{}}\toprule
65-suite channel, $k=3$ & Releases & $b=0$ & $b=16/25$ & $b=1$\\\midrule
Stock catalogue &65&$38/39$&$38/39$&$38/39$\\
Standards-aware &59&$62/65$&$62/65$&$62/65$\\
Within joint fibres &86&$20/39$&$20/39$&$20/39$\\\bottomrule
\end{tabular}
\end{center}
There are 13 singleton fibres, nine pairs, four triples, two of size five and two of size six. Substitution into (\ref{eq:fibres}) gives $20/39$. In the first two stored codebooks a suitable first question already achieves their full-budget score. Equality of the endpoints and monotonicity explain the constant columns. Only the last row at full budget is certified globally optimal by (\ref{eq:fibres}).

\paragraph{Reproduction and independent checks.} The program enumerates every deterministic first question and every subset of its answer branches on which to pay. Each local action contributes its joint probability of spending and of guessing correctly. Taking the upper convex hull permits randomized choices. Adding local curves in decreasing order of their slopes allocates the single global budget. An independent sequence-form linear program uses first-action probabilities $z_{ri}$ and paid-branch probabilities $0\le y_{riD}\le z_{ri}$ with $\sum_i z_{ri}=1$. Its cost constraint is $\sum_{riD}y_{riD}\Pr(R=r,X\in D)\le b$, where each $D$ is a first-answer fibre. Both methods agree on all reported test cases. The exact tests cover grids of sizes $2\times3$, $3\times3$ and $4\times7$, seven budgets, all private sampling tickets at the tested settings, the two recalibrated populations and the frozen 65-suite catalogue. The results file records exact rational probabilities. For irrational $s_b$ the sampler uses a rational setting $s'$ on a chosen mesh. In the interior uncapped regime its excess success probability is $(s'-\sqrt b)^2/s'$.

\section{The geometry of a menu of binary audit questions}\label{sec:codes}
A policy designer can change the menu of permitted questions while keeping the set of events and the recorder's private capacity fixed. Consider a uniform event in a binary linear code $C\le\mathbb F_2^\ell$ of dimension $d\ge2$. The menu has one truthful question per coordinate. The reader chooses its first coordinate after the receipt and may choose \emph{any} remaining coordinate after the first answer, subject to the same expected follow-up budget $b$. The recorder still has three private roles. The event to recover is the entire codeword. Write $M_b(C)$ for the optimal probability of identification in this game. Bit queries are atomic. Authorizing arbitrary linear combinations would change the menu.

\paragraph{Theorem 9 (classification of the query menu).} Let
\[
 t(C)=\min_{\substack{u,v\in C\setminus\{0\}\\u\ne v}}
       |\operatorname{supp}u\cap\operatorname{supp}v|.
\]
For every such code its full minimax curve is
\begin{equation}
 M_b(C)=
 \begin{cases}
 M_b\ \text{of (\ref{eq:sharp-feedback})},&t(C)=0,\\
 \min(1,\,2/3+b/2),&t(C)>0.
 \end{cases}\label{eq:code-class}
\end{equation}
Support intersection is a classical coding-theoretic property [19]. The assertion is its equivalence to a complete optimal-disclosure curve when a receipt must be decoded with three private states. The two curves first depart above $4/9$. In an intersecting code a reader can guarantee perfect identification at $b=2/3$ for \emph{every} admissible encoder. With a disjoint pair, for each $b<1$ an encoder exists that keeps the reader's success strictly below one.

\paragraph{Proof.} The converse in Theorem~7 only needs to separate the two most likely atoms with one question and distinguish the remaining pair with a second. Binary coordinates have these properties on a code because distinct words differ somewhere. Suppose $u,v$ have disjoint supports. Choose a public translate of the triple $(0,u,v)$ and assign its posterior $(2s-1,1-s,1-s)$ in that order. The uniform mixture over all translates is Bayes plausible. Every informative coordinate isolates one of the two arms. No coordinate isolates the pivot. The remaining pair can be distinguished by a coordinate on the other arm's support. A constant first coordinate gives no information and even a paid second coordinate then performs at most as well as one informative free query. Thus the attainable curve for this channel is exactly $h_b(s)$. Minimizing over $s$ proves the first line.

Now suppose $t(C)>0$. In any triple $\{x,y,z\}$ the two nonzero differences $y+x$ and $z+x$ meet in some coordinate. There $y$ and $z$ agree and $x$ differs, so the reader can isolate \emph{any designated word} of that triple. With a posterior $p_1\ge p_2\ge p_3$ isolate the heaviest atom. The free question succeeds with probability $p_1+p_2\ge2/3$. The ambiguous branch has mass $1-p_1\le2/3$. Receipts with fewer than three candidates are already resolved by a coordinate. Aggregate free success $s$ and ambiguous mass $U$ therefore satisfy $s\ge2/3$, $U\le2/3$. Paying on a common fraction of ambiguous branches gives success at least
\[
 s+(1-s)\min(1,b/U)\ \ge\ \min(1,2/3+b/2).
\]
For equality use uniform posteriors on the translates of any triple. Every informative first bit leaves a singleton and an equiprobable pair. The ambiguous mass is $2/3$ and each paid ambiguity earns exactly $1/2$ per query. This proves the second line, with a single encoder valid at all budgets.\hfill$\square$

\paragraph{Corollary 2 (sharp number of withdrawals).} In an intersecting code, $t(C)$ is the least number of coordinate questions that can be withdrawn to move the menu to the other curve \emph{while preserving all $2^d$ distinct event identities}. Deleting fewer coordinates leaves a common position in every pair of nonzero words. For sufficiency take a minimizing pair $u,v$ and delete exactly their overlap. The projection still has full rank. Otherwise some nonzero $w$ would have support contained in the deleted overlap and hence contained in $\operatorname{supp}u$. Proper containment makes $w$ and $u+w$ disjoint, impossible in an intersecting code. Equality forces $w=u$ and $\operatorname{supp}u\subseteq\operatorname{supp}v$, again impossible since $u\ne v$. The projected pair is disjoint and nonzero. Theorem~9 applies. For a full-row-rank generator $G$ with columns $g_j$, the operational threshold can be computed without searching all submenus:
\[
 t(C)=\min_{a,b\ \mathrm{independent}}
        \#\{j:a\cdot g_j=b\cdot g_j=1\}.
\]
For the binary simplex menu containing each nonzero $g_j\in\mathbb F_2^d$ once, the simultaneous equations have $2^{d-2}$ solutions, all nonzero. The required number of withdrawals thus grows as $2^{d-2}$, although the same three private roles and two adaptive queries suffice throughout.

\paragraph{Exhaustive menu experiment.} The dimension-three simplex menu consists of seven parity questions on eight equiprobable events. Enumerate every subset of the seven columns, keep exactly those of rank three, construct the appropriate translation encoder, and evaluate every first bit and the best second bit separately on each answer branch with exact rational arithmetic. At $b=16/25$ the results are
\begin{center}\small
\begin{tabular}{@{}rrrrr@{}}\toprule
Retained questions & Full-rank menus & $t(C)$ & Optimal success & Event count\\\midrule
3 &28&0&$24/25$&8\\
4 &35&0&$24/25$&8\\
5 &21&0&$24/25$&8\\
6 &7&1&$74/75$&8\\
7 &1&2&$74/75$&8\\\bottomrule
\end{tabular}
\end{center}
These are 92 labelled menus. Removing one question from the full seven preserves the curve. Removing an appropriate pair changes it by exactly $2/75$ at this budget. For simplex dimensions two through five the measured withdrawal thresholds are $1,2,4,8$, respectively. The checks verify recovery for every event in all 92 menus. These counts are design-space enumerations. They give no estimate of how often a deployed reviewer requests a particular bit.

\subsection{A fourth private role changes the controlling property}
Write $M_b^{(4)}(C)$ for the same game with four distinguishable private states. Call a quartet balanced by a coordinate when exactly two of its four words have bit one there. Balancing every quartet is the $(|C|,4,2)$ splitter condition of [21]. It gives a different operational criterion.

\paragraph{Theorem 10 (the full four-role curve).} For every uniform binary linear event code $C$ of dimension at least two and every expected follow-up budget $0\le b\le1$,
\begin{equation}
 M_b^{(4)}(C)=
 \begin{cases}
 (1+b)/2,&\text{each quartet admits a balanced coordinate},\\[2pt]
 H(b):=\min\{1/2+b/3,\;3/4\},&\text{some quartet has no balanced coordinate}.
 \end{cases}\label{eq:quartet-full}
\end{equation}
A single recorder in each case attains its entire curve. At $0<b<3/4$, for an arbitrary full-support prior on a finite set of distinct binary words with at least four elements, equality with $H(b)$ holds exactly when the prior is a mixture of uniform distributions on unbalanced quartets. The mixing coefficients compile directly into an exactly decodable channel by Bayes inversion.

\paragraph{Proof.} When a balancing question exists on every quartet, extend any smaller support to a quartet and ask its balancing bit. Each branch now contains at most two candidates. A first guess succeeds with probability at least $1/2$ and a second bit identifies the event. Paying independently with probability $b$ gives $(1+b)/2$. Uniform translates of any two-dimensional linear subspace attain equality. Every informative first bit on this law leaves two equiprobable pairs. A paid question can correct at most half an error per query.

The unbalanced case requires controlling a reader that spends its quota preferentially after particular receipts and answers. Price a paid query at $\lambda\in[0,1]$ and let $B_\lambda(p)$ be the maximum correct-guess probability minus the expected query charge at posterior $p$. The finite decision-tree duality of (\ref{eq:price-envelope}) applies to this menu and four roles. In particular for any fixed channel
\[
 V_b^\pi(W)=\min_{0\le\lambda\le1}
  D_W(\lambda),\qquad D_W(\lambda)=\lambda b+\mathbb E_R B_\lambda(p_R).
\]
We claim that every admissible posterior satisfies
\begin{equation}
 B_\lambda(p)\ge g(\lambda):=\max\{1/2,\,3(1-\lambda)/4\}.\label{eq:quartet-price}
\end{equation}
One free question can separate the two heaviest words. Their total mass is at least $1/2$, strictly larger unless the posterior is uniform on four words. On a balanced quartet, spending on both branches gives $1-\lambda$. For an unbalanced quartet every informative first bit isolates one word. At least three words can be isolated: otherwise the two remaining words would agree at every coordinate. Choose an isolatable word of greatest mass $t$. A globally least atom of mass $m\le1/4$ can be left in the other branch, and $t\ge m$. On that triple choose a second bit separating its two heaviest atoms. Paying only on the triple gives success $1-m$ and cost $1-t$. Its priced value, minus $3(1-\lambda)/4$, is
\[
 (1-\lambda)(1/4-m)+\lambda(t-m)\ge0 .
\]
This also covers smaller supports by padding with zero masses. On a uniform unbalanced quartet the best free success is $1/2$. Paying after its ambiguous answer increases success by $1/4$ for a cost of $3/4$. A constant first bit, or payment on a singleton, yields no higher value. This gives equality in (\ref{eq:quartet-price}) at every price. Minimizing $\lambda b+g(\lambda)$ yields $H(b)$; uniform translates of the quartet realize the uniform prior and preserve the branch law.

For $0<b<3/4$ the minimizing price is uniquely $1/3$. Equality in the channel bound requires equality at this price for every receipt of positive probability. A nonuniform posterior has free success greater than $1/2$ by the two-heaviest argument, and a balanced quartet has priced success at least $2/3$. Only the stated uniform unbalanced posteriors remain. Their barycenter is necessarily $\pi$; conversely any such barycenter realizes the curve. Attainment for finite alphabets follows from the compact posterior-mixture argument in Section~\ref{sec:feedback}, now on faces of size at most four. Finally, every intersecting code balances every quartet: translate the quartet to $\{0,u,v,w\}$ and take a common support position of $u+v$ and $w$. There exactly one of $u,v$ and also $w$ have value one.\hfill$\square$

\paragraph{What the criterion decides.} The menus $\{3,5,6,7\}$ and $\{4,5,6,7\}$ are indexed by three-bit masks, retain eight event identities, and have identical three-role curves. With four roles the first balances every quartet. In the second the output words $\{0,10,12,9\}$ witness the unbalanced case, counting the bit for column $4$ as least significant. At a target correct-guess probability of $7/10$ the maximum admissible budgets are respectively $2/5$ and $3/5$. In the 92 retained menus the two cases occur 57 and 35 times. We checked both constructions by direct enumeration of transcript decisions at six budgets. For all 6,440 menu/quartet pairs we also tested the local price inequality at five prices and all 35 strictly positive posterior weight vectors of denominator eight. Identical induced bit partitions were evaluated once, covering 1,127,000 indexed comparisons. These tests do not substitute for the all-budget proof.

\paragraph{Compiling a prior on three binary audit fields.} For the three coordinate questions on $X=\mathbb F_2^3$, the unbalanced quartets are exactly the eight translates of the Hamming ball $T=\{0,e_1,e_2,e_3\}$. To see this, each useful coordinate isolates one word and three different words must be isolatable. The remaining word has the majority value in each coordinate. The other words flip one coordinate each. Theorem~10 now turns an attainability question into a receiver-independent calibration test. Let $\chi_S(x)=(-1)^{\sum_{i\in S}x_i}$ and $a_S=\mathbb E_\pi\chi_S(X)$. For one representative $z$ of each antipodal pair $\{z,\bar z\}$, where $\bar z=z+(1,1,1)$, define
\[
 L(z)=2\sum_{i=1}^3 a_{\{i\}}\chi_{\{i\}}(z)-2a_{123}\chi_{123}(z).
\]
For every interior $0<b<3/4$ the exact optimal curve equals $H(b)$ if and only if
\begin{equation}
 a_{12}=a_{13}=a_{23}=0,\qquad
 \sum_{\{z,\bar z\}} |L(z)|\le4 .\label{eq:cube-test}
\end{equation}
Indeed, if $\omega$ is the law of the published ball centre then $\pi=\omega*{\rm Unif}(T)$. In the Boolean character basis [22] the multiplier is $(1+\sum_i(-1)^{\mathbf1_{i\in S}})/4=1-|S|/2$. Degree-two coefficients therefore vanish. Fourier inversion fixes the odd part of $8\omega(z)$ to $L(z)$ and leaves the degree-two coefficients free. The even part can take four arbitrary pair values $E_z$ of mean one: the constant and the three degree-two characters span this four-dimensional space. Both members of a pair have nonnegative probability precisely when $E_z\ge|L(z)|$. This proves necessity and sufficiency. An explicit compiler takes $d=1-\sum_{\rm pairs}|L|/4$ and $\omega(z)=(|L(z)|+d+L(z))/8$. It emits centre $z$ on input $x\in z+T$ with probability $\omega(z)/(4\pi(x))$ and retains the two-bit rank of $x$. A marginal bias alone can break calibration: for $\pi(x)=(1+\eta\chi_1(x))/8$ the exact admissibility range is $|\eta|\le1/2$, although all three degree-two tests vanish throughout. On the grid of all 6,435 positive count vectors summing to 16, 81 satisfy (\ref{eq:cube-test}). An independent feasibility LP agreed on 255 selected vectors, including all 81 accepted ones.

There is also a way to recover this spectral calculation from the decision rules themselves. At the generated interior price $\lambda=1/3$, the symbolic experiment determines the value of each uniform quartet from its tree profiles. Exactly eight quartets attain $1/2$. The program forms their normalized event-incidence columns, labels each by its coordinatewise majority and diagonalizes the resulting matrix. Its eigenvalues are $1,1/2,0,-1/2$ with multiplicities $1,3,3,1$. Three zero modes account for the three parity equalities in (\ref{eq:cube-test}); positivity on the remaining modes accounts for the antipodal inequality. Across all 92 full-rank menus at capacities three and four, the compiler solved 18 local split systems and reproduced all 184 complete budget profiles. The implementation receives the bit masks and capacity; the independent comparison script contains the predicted curves.

\paragraph{The price of a two-field correlation.} Let $\epsilon=\mathbb E_\pi\chi_{ij}$, $e=|\epsilon|$, and $u_+=\max(u,0)$. For any full-support law on the cube the following expression bounds the optimal identification probability from below. It is the \emph{exact} optimum for the entire one-parameter family $\pi_\epsilon(x)=(1+\epsilon\chi_{ij}(x))/8$, for every $0\le b\le1$:
\begin{equation}
 J_e(b)=\frac12+\frac12\min(b,e)
          +\frac13\min\big((b-e)_+,\;3(1-e)/4\big).
 \label{eq:drift-gap}
\end{equation}
In particular, unchanged individual bit frequencies say nothing about the persistence of the optimum when their joint distribution changes.

To prove the lower bound, set $g_P(\lambda)=\max(1/2,1-\lambda)$ and $h=|\mathbb E_p\chi_{ij}|$ for one receipt posterior. We claim the price interpolation
\[
 B_\lambda(p)\ \ge\ (1-h)g(\lambda)+h g_P(\lambda).
\]
On any balanced quartet, $B_\lambda\ge g_P$ by the balancing question. On smaller supports one question separates the two heaviest atoms and two adaptive bits identify all candidates, again giving this bound. On an unbalanced ball two cells have each parity. Its two heaviest atoms therefore carry mass at least $(1+h)/2$. Its lightest atom has mass $m\le(1-h)/4$, by taking the smaller parity class. For $\lambda\le1/3$ the paid-on-triple policy in Theorem~10 has success at least $(3+h)/4$ and its priced value at $\lambda=1$ is nonnegative. Affine interpolation yields $B_\lambda(p)\ge(1-\lambda)(3+h)/4$, as required. For $\lambda\ge1/3$ the free success $(1+h)/2$ suffices. Taking expectations gives the claimed interpolation with $e$ since $\mathbb E h\ge e$ and $g_P\ge g$. Minimizing over price now gives (\ref{eq:drift-gap}): the only possible breakpoints of the priced bound are $1/3$ and $1/2$, together with the endpoints.

For attainability under $\pi_\epsilon$, use uniform ball posteriors of total weight $1-e$ with all eight centres equally represented. Assign the remaining weight $e$ to the uniform posterior on the four words of the more frequent parity. Their barycenter is $\pi_\epsilon$, so Bayes inversion implements these exact posterior weights. On the published parity plane any informative coordinate leaves two equiprobable pairs. Its priced value is $g_P$. The average priced value of the recorder is therefore exactly the interpolation above. Operationally, the reader spends first on plane receipts at gain $1/2$ per query, then on ambiguous ball answers at gain $1/3$, until the quota $e+3(1-e)/4$ is exhausted. The same recorder is optimal at every budget. The transcript program agrees with this curve for 19 signed tilts, each of the three pairwise parities, and 21 budgets (1,197 evaluations). Releasing the plane identifies parity on that receipt; this application permits disclosure of parity and protects the exact suite.

\paragraph{Audit decision on negotiated TLS sessions.} Operational TLS migration logs can expose the negotiated cipher name alongside other request attributes [23]. A recorder servicing a restricted review API can retain exact suite recovery while giving an ordinary reader the right to query three fields: DHE versus static RSA, GCM versus CBC, and AES-256 versus AES-128. We selected the complete eight-cell factorial subset of RSA-authenticated suites from the 65-suite catalogue, using the SHA-256 CBC variants. The observer in this test receives only a receipt and the permitted answers. A reader of the handshake itself, or of a second log containing the cipher name, already knows the suite. The policy is meaningful for a separate audit archive and for the single-event observer defined here.

We ran 960 fresh TLS~1.2 handshakes between two local TCP endpoints using Python's TLS binding to OpenSSL 3.5.8. The client verified a temporary local certificate, offered one selected suite per connection and checked the negotiated name. The server read the negotiated suite from its TLS session, sampled the receipt and wrote a one-byte public identifier and a separately packed two-bit private role. We reopened both files and compared the recovered sequence with the complete run schedule. All 960 suite identities matched. Each population used 320 public bytes and 80 private bytes, excluding public configuration and filesystem overhead. Three prescribed populations were independently exercised:
\begin{center}\small
\begin{tabular}{@{}lrrrl@{}}\toprule
Session counts per suite & Recorder & $V_{2/5}$ & $V_{3/5}$ & Target $7/10$\\\midrule
40 each & ball & $19/30$ & $7/10$ & $b=3/5$ feasible\\
50 or 30 by DHE bit & calibrated ball & $19/30$ & $7/10$ & $b=3/5$ feasible\\
60 or 20 by DHE/GCM parity & ball and plane & $7/10$ & $47/60$ & $b\le2/5$ required\\\bottomrule
\end{tabular}
\end{center}
The table's identification probabilities are exact best responses computed from the recorders. The sample counts and recovery checks are measurements. In the last population the marginals of all three bits remain balanced but $a_{12}=1/2$. Equation~(\ref{eq:drift-gap}) gives $J_{1/2}(3/5)=47/60$, exceeding the $7/10$ ceiling even for the best possible recorder. Solving $J_{1/2}(b)\le7/10$ gives $b\le2/5$ exactly. With the public centre or plane identifier and the same private rank size, the admissible rate falls from 60 to 40 follow-ups per 100 sessions in expectation. A raw suite-name archive gives the reader success one before asking a question. The protected mechanism still supplies exact reconstruction for every session. The local trials have prescribed population laws and exclude other telemetry. Field estimation, linkage across sessions, and enforcement of query quotas are separate system requirements.

\section{Scope, prior results and tests of the model}
\lead{Where the secrecy resides} The private index must remain available and secret at audit time. If the archive observer obtains it, exact recovery makes the fact public to that observer. If the index is lost, exact recovery fails. An additional protected key or a separate store changes the capacity accounting. We assume an honest recorder and an authentic later label. Sections~\ref{sec:feedback} and \ref{sec:codes} analyze receipt-dependent two-stage policies. Forged records, computationally bounded observers and noisy reconstruction require further security experiments.
\par\lead{Relation to prior work} The guessing metric is established [1,2]. Hard admissibility constraints on privacy mechanisms [3] and graph-theoretic zero-error leakage [4] supply close precedents. The local cover objective is related to fractional local coloring [5]. Here supports are restricted by private recovery capacity and contexts are specified independently of those supports. Causal secure source coding [6] studies rate and secrecy under timing constraints. Database auditing under retention policies [10] studies which queries remain answerable after history is suppressed. Our quantity is a finite, worst-context guessing factor when the timing of public side information changes at fixed recovery capacity. AgentSecBench measures security games under permitted projections [7]. Observer-Quotient Security bounds composition defects [8]. Neither role supplies the extremal separation proved here. Channel design with an optimizing receiver also has established game-theoretic formulations [20]. Cost-sensitive diagnostic policies [24] minimize measurement plus misclassification cost, equivalent here to maximizing correct guesses minus a test price for a fixed posterior. Our additional minimization ranges over every zero-error recorder at a fixed private capacity. Its four-role solution (\ref{eq:quartet-full}), the population test (\ref{eq:cube-test}) and the exact parity-tilt response (\ref{eq:drift-gap}) follow from constraints on the posteriors that the recorder can realize. Earlier in the paper the budget-dependent posterior weights, the equality condition (\ref{eq:corner-prior}) and the menu classification (\ref{eq:code-class}) are derived specifically for the present receipt game. A priority claim under other formulations would require a broader review.
\par\lead{An operational test} In an audit system with timestamped policy decisions, determine whether the policy class was available to the recorder before receipt generation, identify every protected state accessible to the decoder, and enumerate admissible event/label pairs from the policy engine. The exogenous guessing factor needs only the incidence relation. In the adaptive experiment the inference score also requires a population law, as (\ref{eq:drift-gap}) illustrates. Compute both optima on held-out policy versions and compare them to the implemented channel. A public guessing factor below the certified lower bound with exact recovery would falsify the mapping: check for an extra private input, an unavailable context, a receipt-dependent label, or a missing public observation. A gap that vanishes throughout the held-out workload would limit the practical reach of the separation.
\par\lead{Origin and limits of the rate results} Brakensiek [11] supplies the sharp power inequality for clique neighborhoods; Alon and Defant [12] supply the finite profile recursion. Theorem~4 translates that graph theory into an audit leakage rate. Theorem~5 also covers multiple context orbits and yields the grid calculation. Its proof uses a weighted neighborhood bound and a public permutation to control the worst sequence of label types. Establishing the prior-art boundary for equivalent information-theoretic formulations, and measuring a real recorder under an observed policy schedule, are separate research tasks.

\end{document}